# CP Nonconservation in $p\bar{p} \to t\bar{b}X$ at the Tevatron

David Atwood[a], Shaouly Bar-Shalom[b], Gad Eilam[b] and Amarjit Soni[c]

a) Theory Group, CEBAF, Newport News, VA 23606, USA

b) Department of Physics, Technion, Haifa, Israel

c) Department of Physics, Brookhaven National Laboratory, Upton, NY 11973, USA

**Abstract**: The reaction $p\bar{p} \to t\bar{b}X$ is found to be rather rich in exhibiting several different types of CP asymmetries. The spin of the top quark plays an important role. Asymmetries are related to form factors arising from radiative corrections of the $tbW$ production vertex due to non-standard physics. As illustrations, effects are studied in two Higgs Doublet Models and in Supersymmetric Models; asymmetries up to a few % may be possible.

The origin of CP violation remains a pressing issue in Particle Physics. The Standard Model (SM), with three generations of quarks, can accommodate a CP violating phase, the Cabibbo-Kobayashi-Maskawa (CKM) phase [1]. However it is widely believed that this phase cannot account for baryogenesis [2]. Additional CP violating phases due to new physics are therefore a necessity. Besides, in extensions of the SM, new phase(s) appear rather readily. It is therefore quite unlikely that the CKM phase is the only CP violating phase in nature. In particular, in top physics the SM causes negligible CP violation effects [3] whereas, in sharp contrast, non-standard sources often give rise to appreciable effects [4, 5]. Searching for CP violation in top-quark production and/or decay is therefore one of the best ways to look for signals for new physics.

In this paper we examine CP violation asymmetries in top quark production via the basic quark-level reaction:

$$u + \bar{d} \to t + \bar{b}, \quad \bar{u} + d \to \bar{t} + b \tag{1}$$

Indeed the reaction is very rich for CP violation studies as it exhibits many different types of asymmetries. Some of these involve the top spin. Therefore the ability to track the top spin through its decays becomes important and top decays have to be examined as well.

It is easy to see that in the SM these effects are expected to be extremely small as they are severely suppressed by the GIM mechanism [3]. As illustration of the possibilities with non-standard sources of CP violation we consider two extensions of the SM: a two Higgs Doublet Model (2HDM) with Natural Flavor Conservation (NFC), often called type II model and a Supersymmetric Standard model (SSM). We find that CP asymmetries can be sizable, in some cases at the level of a few percent. Thus the asymmetries in $t\bar{b}$ production can be appreciably larger than those in $t\bar{t}$ pair production [6, 7] wherein they tend to be about a few tenths of percents. Since the number of events needed for observation scales as (asymmetry)$^{-2}$ the enhanced CP violation effects in $t\bar{b}(\bar{t}b)$ may make up for the reduced production rates for $t\bar{b}$ compared to $t\bar{t}$. In fact larger asymmetries are not just gratifying but they can also be essential as detector systematics can be a serious limitation for asymmetries $\lesssim 0.1\%$.

Let us first discuss the asymmetries in the $u\bar{d}$ ($\bar{u}d$) subprocess. We consider four types of asymmetries that may be present. First is the CP violating asymmetry in the cross section:

$$A_0 = (\sigma_q - \sigma_{\bar{q}})/(\sigma_q + \sigma_{\bar{q}}) \tag{2}$$

where $\sigma_q$ and $\sigma_{\bar{q}}$ are the cross-sections for $u\bar{d} \to t\bar{b}$ and $\bar{u}d \to \bar{t}b$ respectively at $\hat{s} = (p_t + p_{\bar{b}})^2$. The CPT theorem of quantum field theory implies that the total cross-section for $u\bar{d}$ and $d\bar{u}$ are identical. If a cross-section asymmetry $A_0$ in the $t\bar{b}$ final state is present, then to maintain the balance of total cross sections, another mode must have a compensating asymmetry.

The spin of the top allows three additional types of CP violating polarization asymmetries. To define these let us introduce the co-ordinate system in the *top-quark rest frame* where the unit vectors are given by $\vec{e}_z \propto -\vec{P}_b$, $\vec{e}_y \propto \vec{P}_u \times \vec{P}_b$ and $\vec{e}_x = \vec{e}_y \times \vec{e}_z$. Here $\vec{P}_b$ and $\vec{P}_u$ are the 3-momenta of the $\bar{b}$ quark and the initial $u$-quark in that frame. We denote the longitudinal polarization or helicity asymmetry as

$$A(\hat{z}) = (N_R - N_L - \bar{N}_L + \bar{N}_R)/(N_R + N_L + \bar{N}_L + \bar{N}_R) \quad (3)$$

where $N_L$ is the number of left-handed top quarks produced in $u\bar{d} \to t\bar{b}$ and $\bar{N}_R$ is the number of right-handed $\bar{t}$ produced in $\bar{u}d \to \bar{t}b$ etc. In the frame introduced above therefore right handed tops have spin up along the $z$-axis and left handed ones spin down. We further define the CP violating spin asymmetries in the $x$ and $y$ directions as follows:

$$A(\hat{x}) = (N_{x+} - N_{x-} + \bar{N}_{x+} - \bar{N}_{x-})/(N_{x+} + N_{x-} + \bar{N}_{x+} + \bar{N}_{x-}) \quad (4)$$
$$A(\hat{y}) = (N_{y+} - N_{y-} - \bar{N}_{y+} + \bar{N}_{y-})/(N_{y+} + N_{y-} + \bar{N}_{y+} + \bar{N}_{y-}) \quad (5)$$

where $N_{j+}(\bar{N}_{j+})$ represent the number of $t(\bar{t})$ with spin up with respect to $\hat{j}$ axis, for $j = x, y$.

While all these four asymmetries are manifestly CP-violating, $A_0$, $A(\hat{z})$ and $A(\hat{x})$ are even under naive time reversal ($T_N$) whereas $A(\hat{y})$ is $T_N$-odd. So the first three require a complex Feynman amplitude whereas $A(\hat{y})$ needs a real amplitude. Of course, all four do need a CP-violating phase in the underlying theory.

In the limit of massless $u$ and $d$ quarks the CP violating contribution to the $Wtb$ vertex may be represented by the effective interaction:

$$\begin{aligned}\mathcal{L} &= i2^{-\frac{1}{2}} g_W \; W^+_\mu \; \bar{t} \left[ F\gamma^\mu + im_t^{-1} G\sigma^{\alpha\mu} q_\alpha \right] L \; b \\ &- i2^{-\frac{1}{2}} g_W \; W^-_\mu \; \bar{b} \; R \left[ F\gamma^\mu + im_t^{-1} G\sigma^{\alpha\mu} q_\alpha \right] t \end{aligned} \quad (6)$$

where $L = (1 - \gamma_5)/2$, $R = (1 + \gamma_5)/2$, we have taken the b-quark to be massless and consider only the left handed component which will interfere with the standard model (tree-level) contribution. The asymmetries of interest are therefore expressible in terms of $F$ and $G$ alone. We will denote the real and imaginary parts by $F = F_R + iF_I$ and $G = G_R + iG_I$. Note that the terms proportional to $F_R$ and $G_R$ are hermitian while the terms proportional to $F_I$ and $G_I$ are not and therefore proportional to final state interaction effects.

We can now derive expressions for the asymmetries in terms of the form factors, for example:

$$A_0 = -2F_I + \frac{6}{2+x}G_I; \quad A(\hat{z}) = 2\frac{2-x}{2+x}F_I - \frac{2}{2+x}G_I \qquad (7)$$

where $x = m_t^2/\hat{s}$. Notice that the three quantities $\{A_0, A(\hat{z}), A(\hat{x})\}$ are linear combinations of the two form factors $\{F_I, G_I\}$, so in particular, one can show that

$$A(\hat{x}) = -3\pi x^{-\frac{1}{2}}\left((2+x)A_0 + (2-x)A(\hat{z})\right)/32 \qquad (8)$$

The dependence of $A_0$ on $F_I$ and $G_I$ provides a clue as to how the balance of total cross-section required by CPT is achieved. In order for these imaginary parts to exist in perturbation theory, there must be a contribution from a loop graph which has an intermediate state $J$ that can be kinematically on shell. $J$ is therefore another component of the cross-section, and in fact it is the cross-section asymmetry in $J$ that compensates for the asymmetry in $t\bar{b}$.

The asymmetry $A(\hat{y})$, on the other hand, is proportional to $G_R$. Since $G_R$ may be related to $G_I$ through a dispersion relation (as a function of $\hat{s}$) we can obtain $A(\hat{y})$ in terms of $A_0$ and $A(\hat{z})$:

$$\begin{aligned}A(\hat{y})[\hat{s}] &= -\frac{3}{32}\left(\frac{1-x}{(2+x)\sqrt{x}}\right)\mathrm{Re}\Bigg(\int_0^\infty \frac{2\xi + x}{(\xi-x)(\xi-1+i\epsilon)\xi} \cdot \\ &\quad \cdot((2\xi - x)A_0[\xi\hat{s}] + (2\xi + x)A(\hat{z})[\xi\hat{s}])d\xi\Bigg)\end{aligned} \qquad (9)$$

Note that the integrand is 0 if $\xi\hat{s}$ is below the threshold to produce an imaginary part. In deriving this relation, we assume that some GIM like cancelation applies in the underlying model in which case all of the asymmetries vanish in the limit of large $\hat{s}$.

Fig. 1a shows the SM tree-level production process. The necessary absorptive parts require radiative corrections, involving a CP violating phase, at least to one-loop order. Fig. 1b shows the only graph relevant to a type-II 2HDM with a CP phase residing in neutral Higgs exchanges. Fig. 1c shows example of a one-loop graph that pertains to a SSM which can involve new CP violating phase(s) as well as the needed absorptive parts.

As is well known, in a 2HDM with NFC, CP nonconservation emanates from soft symmetry breaking complex parameters in the Higgs potential [8, 9]. These induce mixing between real and imaginary parts of the Higgs fields in their mass matrix. Consequently the mass eigenstates do not have a definite CP property. Therefore, an important feature of the 2HDM is that CP-violation may result from the neutral Higgs sector even when there is none in the charged Higgs sector. This CP violating phase from neutral Higgs exchanges is much more difficult to look for compared to that from the charged Higgs exchanges. The top quark can play a special role with regard to the neutral Higgs CP as due to its large mass its coupling with the Higgs are significantly enhanced compared to all the other quarks.

A distinctive manifestation of such CP violation is that the neutral Higgs mass eigenstates couple to fermions with both scalar and pseudoscalar couplings. The relevant Feynman rules for calculating the asymmetries of interest can be extracted from the following part of the Lagrangian,

$$\mathcal{L}_{H_j^0} = H_j^0 \bar{f}(a_{fj} + ib_{fj}\gamma_5)f + c_{Wj} m_W H_j^0 g_{\mu\nu} W^\mu W^\nu \qquad (10)$$

This involves the $\bar{f}fH_j^0$ and the $WWH_j^0$ couplings, where $j = 1, 2, 3$ for the three neutral spin 0 fields. The coupling constants, $a_{fj}$, $b_{fj}$ and $c_{Wj}$ are functions of $\tan\beta$, which is the ratio between the two vacuum expectation values (VEVs) in this model, i.e. $\tan\beta = v_2/v_1$, and of the three mixing angles $\alpha_{1...3}$ which diagonalize the $3 \times 3$ Higgs mass matrix.

For simplicity we assume that two of the three neutral Higgs particles are much heavier compared to the third one. The effects we are seeking are therefore likely to be dominated by the lightest neutral Higgs. We thus omit the index $j$ in Eqn. 10 and denote the couplings of the lightest Higgs with the top and the $W$ as $a_t$, $b_t$ and $c_W$.

From Fig. 1 we see that the imaginary part of the loop is provided by the $WH$ intermediate state and hence the cross-section asymmetry $A_0$ is compensated by an asymmetry in $u\bar{d} \to W^+H$ versus $\bar{u}d \to W^-H$. Clearly this imaginary part can only exist above the $WH$ threshold at $\hat{s} = (m_W + m_H)^2$. Thus below this $A_0$, $A(\hat{z})$ and $A(\hat{x})$ will be identically 0 though $A(\hat{y})$ need not be since it depends only on virtual effects.

Using the Lagrangian (10) the CP asymmetries, $A_0$ and $A(\hat{z})$, resulting from the interference of Fig. 1a and 1b can be readily calculated:

$$A_0 = -\frac{b_t c_W m_W R_0}{16\pi m_t}\left\{(1 - 3y - z)\phi - 2(1-y)(x + xy - xz - 4y)\tau\right\} \quad (11)$$

$$A(\hat{z}) = \frac{b_t c_W m_W R_0}{16\pi m_t(1-x)}\left\{(1 + 3x - 7y - z + 3xy + xz)\phi - 2\tau\left[(x - 2y)^2\right.\right.$$
$$\left.\left. + (3x - 4y)(1 - z + xz) + x(1-x)y(y-z)\right]\right\} \quad (12)$$

where $x = m_t^2/\hat{s}$, $y = m_W^2/\hat{s}$, $z = m_H^2/\hat{s}$, $\phi = \sqrt{1 + y^2 + z^2 - 2y - 2z - 2yz}$, $R_0 = x/(y(2+x)(1-x))$ and

$$\tau = (1-x)^{-1}\tanh^{-1}[(1-x)\phi/(1 + x - y - z + xz - xy)] \quad (13)$$

As should be expected from Eqn. (10) all the CP asymmetries have to be proportional to the product $b_t c_W$. We choose the angles in the Higgs mixing matrix as $\alpha_1 = \alpha_2 = \pi/2$ and $\alpha_3 = 0$ as it tends to give maximal effects[10, 11]. Then one can show that $b_t c_W m_W \simeq .2m_t \cos\beta \cot\beta$ and the asymmetries can now be expressed as a function of $\tan\beta$ and $m_H$ only.

We present our numerical results for $\tan\beta = 0.3$ [12]. Numbers for other values of $\tan\beta$ can then be readily obtained using the relation given in the preceding paragraph. Fig. 2 a and b show the asymmetries as a function of $\hat{s}$ for $m_H = 100$ and 400 GeV. The cross section asymmetry, $A_0$, and helicity asymmetry, $A(\hat{z})$ are typically a few percent. The polarization asymmetry $A(\hat{x})$ likewise is a few percent.

Since the real part of the graph in Fig. 1b does not need a physical threshold, it may receive contributions from Higgs bosons of arbitrary mass. In the limit of degenerate Higgs masses, CP violating effects should vanish. Hence it may not be a legitimate approximation to ignore the more massive Higgs bosons in this loop. We will assume therefore that the other Higgs bosons of the theory have a mass $m'_H$ and, for the purposes of numerical estimates we will take $m'_H = 1TeV$. The effect of these heavy Higgs bosons on the dispersion Eqn. (9) is to replace $\{A_0, A(\hat{z})\}$ by $\{A'_0, A'(\hat{z})\}$ which are defined by subtracting the effects of the heavy Higgs. So, for instance, $A_0$ gets replaced by $A'_0[\hat{s}] = A_0[\hat{s}, m_H] - A_0[\hat{s}, m'_H]$ and likewise for the other asymmetries. Note that if $\hat{s} < (m'_H + m_W)^2$ then $A'_0 = A_0$; $A'(\hat{z}, \hat{x}) = A(\hat{z}, \hat{x})$ since $\hat{s}$ is below the $W$, $H'$ threshold. In Fig. 2a and 2b we also show the value of $A(\hat{y})$. There tends to be a cusp at the $HW$ threshold where the peak value of $A(\hat{y})$ is about one percent.

To search for the effects of these three types of spin asymmetries that occur at the production vertex, decays of the top will obviously need to be examined [13]. In particular when considering $A(\hat{y})$ one must keep in mind that it is dependent on the real part of the loop amplitude of Fig. 1b. One complication that this could lead to is that a similar asymmetry may also enter into the decay $t \to bW$ when similar radiative corrections to that vertex are also included [15]. This is not a concern in the case of the other observables since if we assume that the Higgs is above the threshold, i.e. $(m_W + m_H) > m_t$, the necessary condition that there be an imaginary part in the decay amplitude is not satisfied.

As it turns out, the observed value of $A(\hat{y})$ is not affected by CP violation in the decay process. The key point is that the measurement of $A(\hat{y})$ through the decay chain $u(p_u)\,\bar{d}(p_d) \to \bar{b}(p_{b1})\,t(p_t)$ followed by $t(p_t) \to b(p_{b2})\,e^+(p_e)\,\nu(p_\nu)$ is equivalent to measurement of the term proportional to $\epsilon(p_e, p_d, p_t, p_{b1})$. On the other hand, CP violation arising from the decay process is proportional to $\epsilon(p_e, p_d, p_t, p_{b2})$. It is easy to see that an observable related to the first of these will be insensitive to the second[10].

These quark level asymmetries can be readily converted to the hadron (i.e. $p\bar{p}$) level by folding in the structure functions in the standard manner[16]. The results are shown in Fig. 3 where for each of the asymmetries we apply a cut of $\hat{s} > (m_H + m_W)^2$. At the Tevatron ($E = 2$ TeV) the expected number of events are 900–3000 with an integrated luminosity $3fb^{-1}$–$10fb^{-1}$ respectively[17]. If the collider energy gets upgraded to 4 TeV and/or there are additional luminosity upgrades as have often been discussed, then the number of events can go up by another factor of about 2 to 10[17]. Thus the asymmetries, in the range of a few percent, resulting from some extensions of the SM may well become within the reach of experiment provided, of course, that the signal for these single top events could be extracted from possible backgrounds[18].

Another extension of the standard model which can produce these kind of asymmetries is SSM. There are a number of possible graphs which could contribute[10, 11]; here we will consider only the gluino exchange diagram given in Fig. 1c. In this case CP violation arises through the mixing matrix between the fermion and the scalar states, in general a $6 \times 6$ matrix. For simplicity, let us consider a scenario, motivated by supergravity models, where all the squarks are degenerate with a mass $\tilde{m}_q$ except for the super partner of the top quark, the stop. Furthermore, the two stop states mix with the left and right parts of the top quark with a general $2 \times 2$ unitary mixing matrix $\mathcal{X}$. The cross-section asymmetry therefore is given by:

$$A_0 = 2\alpha_s Im(\mathcal{X}_{11}\mathcal{X}_{12}^*)\left[g(\tilde{m}_{t1}, \tilde{m}_b, \tilde{m}_g, s) - g(\tilde{m}_{t2}, \tilde{m}_b, \tilde{m}_g, s)\right] \qquad (14)$$

where $\tilde{m}_{t1,2}$ are the masses of the two stops. The function $g$ is given by:

$$g = (4/3)\sqrt{xw}(1-x)^{-1}(2+x)^{-1}\phi(Q^{-1}\tanh^{-1}Q - 1) \qquad (15)$$

where $x = m_t^2/s$, $y = \tilde{m}_q^2/s$, $z = \tilde{m}_t^2/s$ and $w = \tilde{m}_g^2/s$ while $\phi$ has the same definitions as previously in terms of $y$ and $z$ and $Q = (1-x)\phi/(1+2w-x-y-z+xz-xy)$

In this case the helicity structure of the model is such that the form factor $F = G$. Thus $A(\hat{z}) = A_0$ from which $A(\hat{x})$ and $A(\hat{y})$ follow through Eqn. (8,9). In Fig. 2 we also show these asymmetries due to the SSM for $\tilde{m}_{t1} = 100$ GeV, $\tilde{m}_{t2} = 500$ GeV, $\tilde{m}_g = 100$ GeV and $\tilde{m}_q = 100$ GeV. We have also assumed that the the quantity $Im(\mathcal{X}_{11}\mathcal{X}_{22}^*) = 1/2$ which is its maximum value. We can see that in this case the asymmetries are less than 1%. The small size of these asymmetries is, in part, due to the fact that the intermediate state (see Fig. 1c) consists of two scalars that must be in a P-wave giving rise to an additional threshold suppression factor. However, in SSM many other types of loop corrections (e.g. box graphs) can also contribute giving rise to asymmetries on the order of several percents. [10, 11].

We close with a few remarks in brief.

1. It is important to note that from the point of view of experimental detection these four asymmetries are independent. Thus, the sensitivity of a given detector to observing the combined CP violation effects may be appreciably better than that for any one asymmetry[10, 11].

2. We have focussed here on a $p\bar{p}$ machine (i.e. the Tevatron) as the self-conjugate nature of the intial state is rather important for CP studies. At the LHC ( i.e. a $pp$ machine), although the event rate is high, such CP studies are quite difficult. Note, for instance, that the cross-sections for $pp \to t\bar{b}X$ and to $\bar{t}bX$ are expected to be different at the LHC even if CP was strictly conserved.

3. We recall that the $W$-glue fusion subprocess, $W^+ +$ gluon $\to t + \bar{b}$, also contributes to the same final state[17]. While it will be useful to include its contribution to the asymmetries in a future study, for now we note that, at least in the 2HDM, CP violating radiative corrections, to one loop order, to $Wg$ fusion do not yield absorptive parts (in the $m_b = 0$ limit).

We will address to some of these issues in greater detail in future work [10].

This research was supported in part by the US-Israel Binational Science Foundation and in part by USDOE contracts DC-AC05-84ER40150 (CEBAF) and DE-AC-76CH00016 (BNL).

[18] We note that experimental issues such as detector efficiencies, backgrounds and the like are clearly very important on their own right, requiring extensive studies that are beyond the scope of this work.

# Figure Captions

Figure 1: Feynman diagrams for contributions to $u\bar{d} \to t\bar{b}$. 1a) the standard model process, 1b) one-loop graph in the two Higgs doublet models (2HDM), and 1c) an example of one-loop graph that could occur in the SSM.

Figure 2: The magnitudes of the quark-level asymmetries $A_0$ (solid); $A(\hat{z})$ (dashed); $A(\hat{x})$ (dotted) and $A(\hat{y})$ (dot-dashed) as a function of $\sqrt{\hat{s}}$ in the (2HDM) with $\tan\beta = 0.3$. Fig. 2a: $m_H = 100$ GeV and Fig. 2b: $m_H = 400$ GeV. Note that $A(\hat{y})$ is computed keeping fixed the masses of the two heavier neutral $H^0$'s to 1 TeV. Also shown with the lower solid line is the asymmetry $A_0$ in the SSM described in the text for parameters $\tilde{m}_{t1} = 100 GeV$, $\tilde{m}_{t2} = 500 GeV$, $\tilde{m}_q = 100 GeV$, $\tilde{m}_g = 100 GeV$ and $Im(\mathcal{X}_{11}\mathcal{X}_{22}^*) = 1/2$.

Figure 3: The corresponding asymmetries in the $p\bar{p}$ c.m., for $E = 2$ TeV, frame as a function of $m_H$ for the 2HDM and as a function of $\Delta\tilde{m}_t (\equiv \tilde{m}_{t2} - \tilde{m}_{t1})$ for the SSM case. See also caption for Fig. 2.

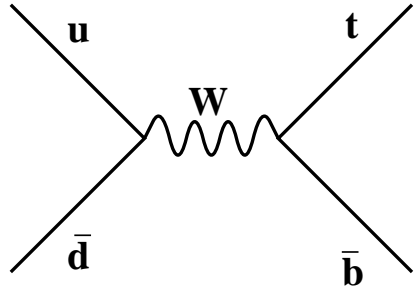

a

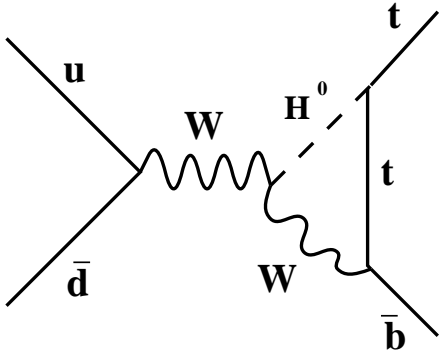

b

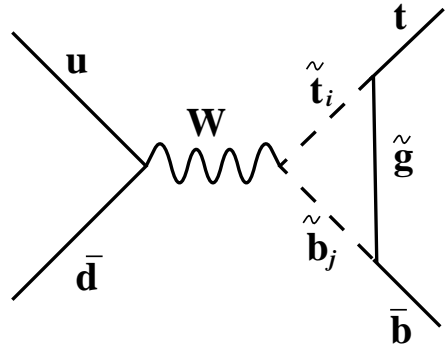

c

**Fig.1**

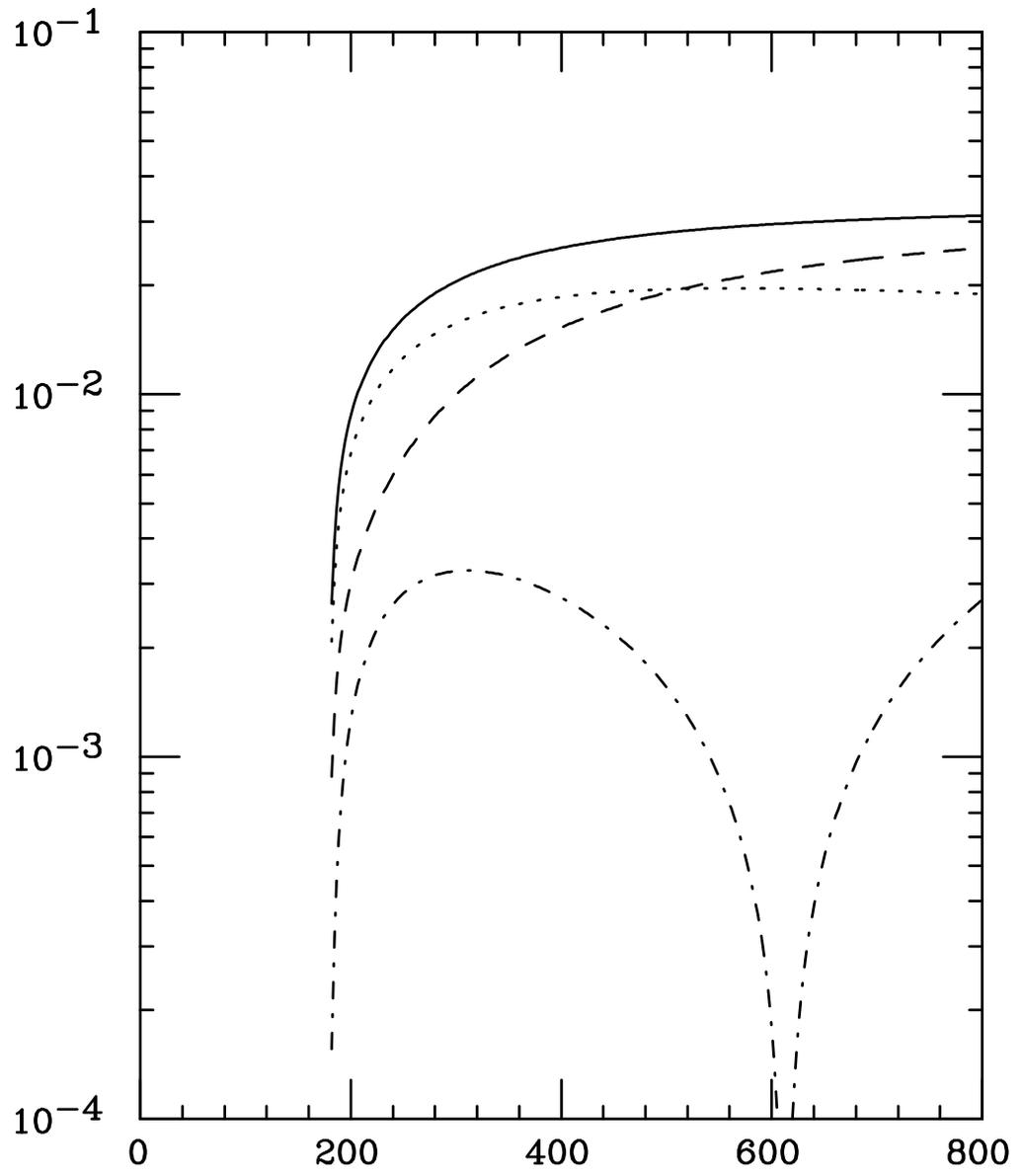

Figure 2a

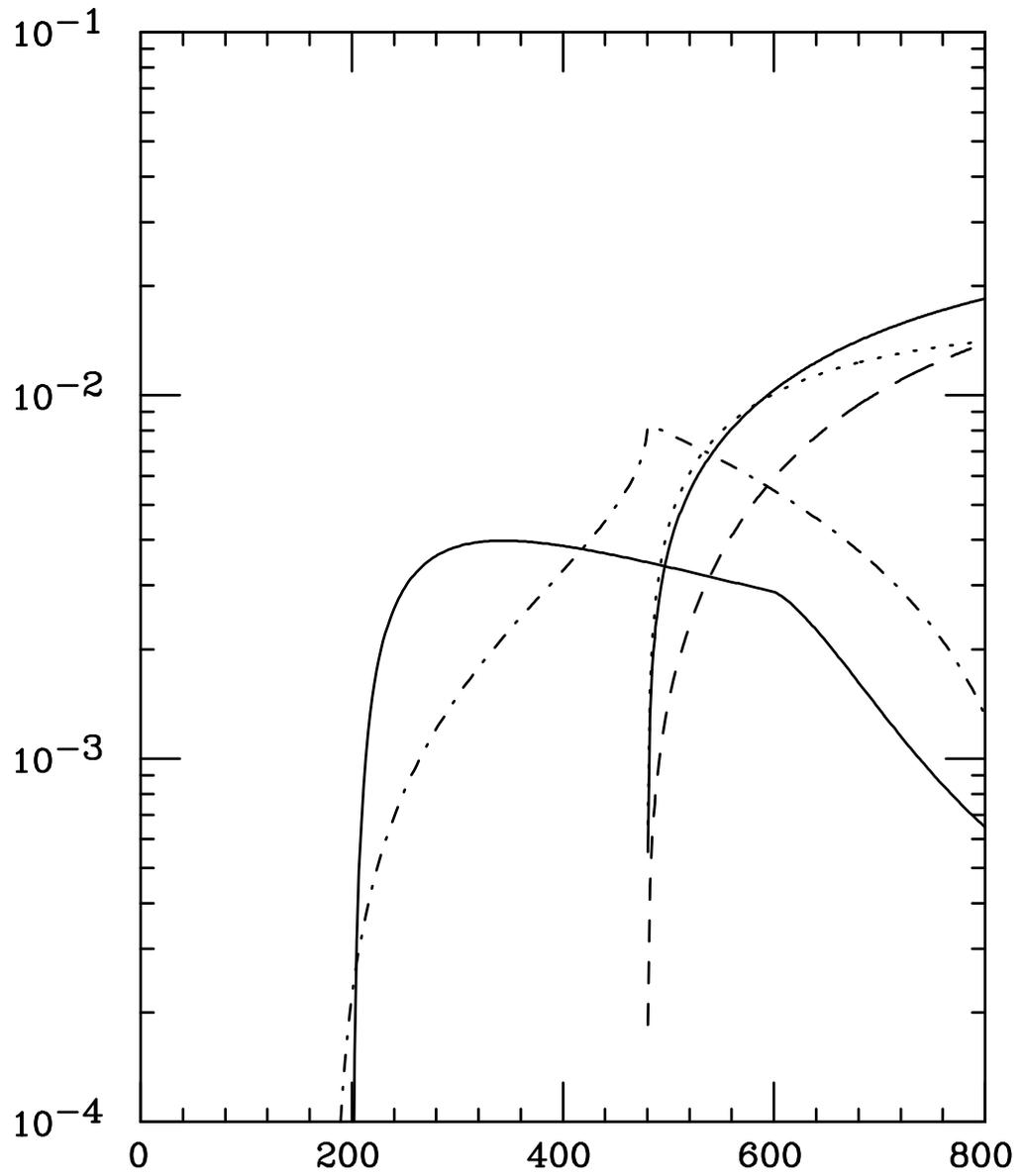

Figure 2b

Figure 3

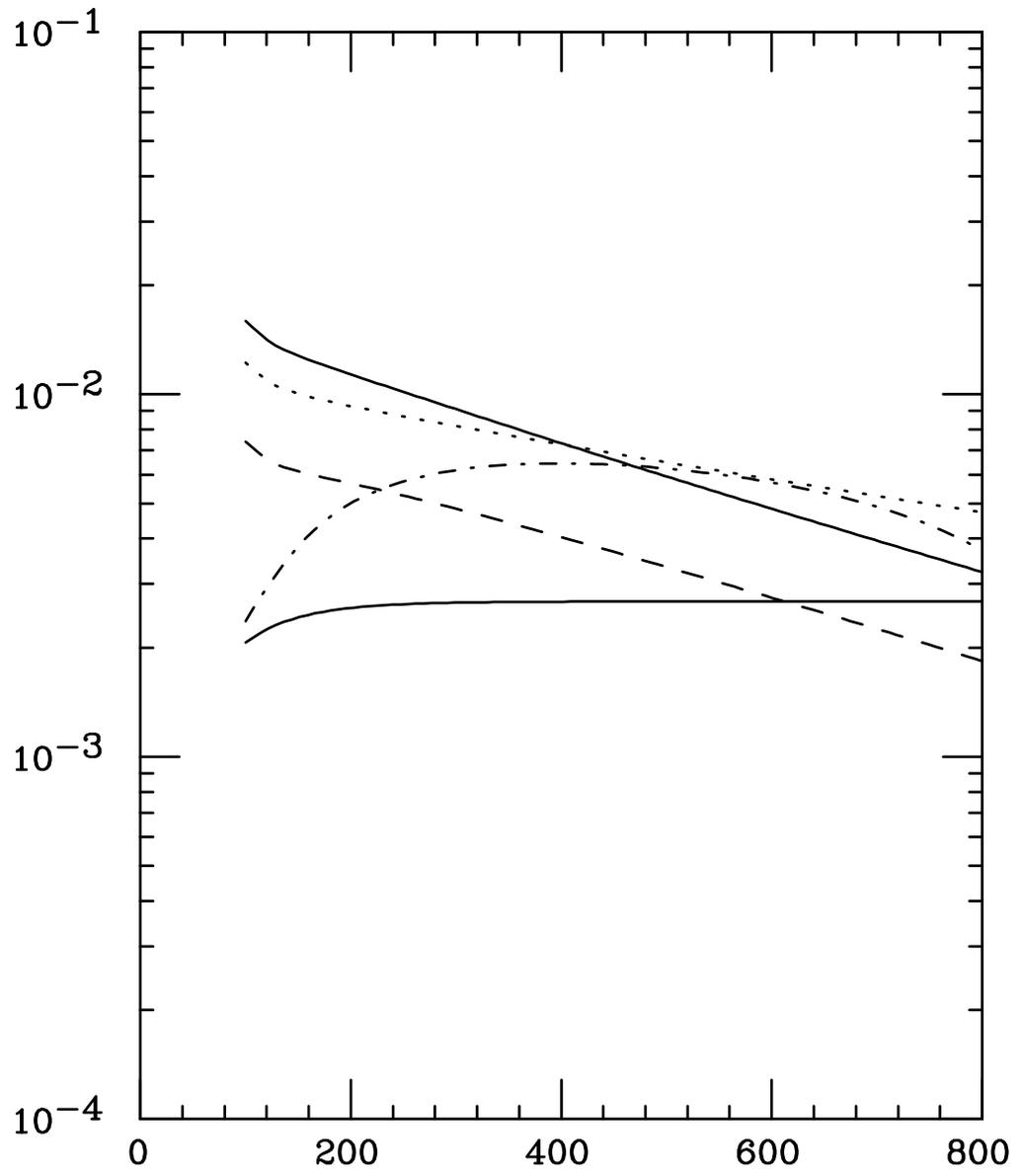